\documentclass[a4paper]{spie}  
\usepackage[]{graphicx}

\title{Development status of a Laue lens project for gamma-ray astronomy} 

\author{F. Frontera\supit{a,d}, G. Loffredo\supit{a}, A. Pisa\supit{a}, L. Milani\supit{a}, 
F. Nobili\supit{a}, N. Auricchio\supit{a}, V. Carassiti\supit{b},
F. Evangelisti\supit{b}, L. Landi\supit{a}, S. Squerzanti\supit{b},
 K.H. Andersen\supit{c}, P. Courtois\supit{c}, L. Amati\supit{d}, E. Caroli\supit{d},
G. Landini\supit{d}, S. Silvestri\supit{d}, J.B. Stephen\supit{d}, J. M. Poulsen\supit{e}, 
B.~Negri \supit{f}, G.~Pareschi \supit{g}
\skiplinehalf
\supit{a}University of Ferrara, Physics Department, Via Saragat 1, 44100
Ferrara, Italy; \\
\supit{b}Istituto Nazionale Fisica Nucleare, Sezione di Ferrara, Via Saragat 1,
44100 Ferrara, Italy; \\
\supit{c}Institute Laue--Langevin, 6 Rue Jules Horowitz, 38042 Grenoble, France\\
\supit{d}INAF, IASF Bologna, Via Gobetti 101, 40129 Bologna, Italy\\
\supit{e}Thales Alenia Italia SpA -- Laben, S.S Padana Superiore, 290, 20090 Vimodrone, Italy\\
\supit{f}Agenzia Spaziale Italiana, Viale Liegi, 26, 00198 Roma, Italy\\
\supit{g}INAF, Osservatorio Astronomico di Brera, 23807 Merate, Italy
}

\authorinfo{Further author information: (Send correspondence 
to F.F.)\\ F.F: E-mail: frontera@fe.infn.it, Telephone: +39 0532 974 254}

\begin{document} 
  \maketitle 

\begin{abstract}
We report the status of the HAXTEL  project, devoted to perform a
design study and the development of a Laue lens prototype.
After a summary of the major results of the design study,
the approach adopted to develop a Demonstration Model of a Laue
lens is discussed, the set up described, and some results presented. 

\end{abstract}


\section{INTRODUCTION}
\label{s:intro}  

The hard X--/gamma--ray astronomy is moving toward a new generation
of telescopes: from direct sky-viewing telescopes to focusing telescopes.
Nowadays it is happening something similar to what happened in the late '70s to 
the soft X--ray astronomy ($<$2 keV) and in the '90s to the 2--10 keV X--ray astronomy,
in the first case with the {\em Einstein satellite}, in the second case with the launch
of the {\em ASCA} and {\em BeppoSAX} satellites, when the first X--ray focusing telescopes
were flown. With the advent of focusing telescopes in the hard X--/gamma--ray astronomy, it
is expected a big leap in sensitivity, by a factor 10--100 with respect
to one of the most sensitive instruments of the current generation, 
and a significant increase in angular resolution (from $\sim 10$ arcmin of the mask 
telescopes like the INTEGRAL IBIS to less than 1 arcmin).

Both the hard X-ray ($<$100keV) and the gamma--ray ($>$ 100 keV) focusing telescope 
generation make use of the Bragg diffraction technique, in the first case, from 
multilayer coatings (ML) in reflection configuration (supermirrors), in the second case, 
from mosaic crystals in transmission configuration (Laue lenses). 

A mission proposal that makes use of both supermirrors and Laue lenses, named 
{\em Gamma Ray Imager}
(GRI), has been recently submitted to ESA in response to the first AO of the 'Cosmic Vision
2015--2025' plan \cite{Knodlseder07}. It covers with unprecedented sensitivity the energy 
band from 10 keV to 1 MeV. While below 100 keV other missions are now under study, 
like Simbol-X \cite{Ferrando06} and NeXT \cite{Takahashi06},
above 100 keV GRI is unprecedented. For the astrophysical importance of the $>$100 keV band 
see, e.g., Refs.~\citenum{Frontera05a,Frontera06,Knodlseder06}. 

Here we report on the current status of our project HAXTEL (= HArd X-ray TELescope) 
devoted to  develop the technology for building
broad energy passband Laue lenses, mainly  for the study of the continuum 
emission of celestial sources above 100 keV.  


\section{Summary of the lens design study results}
Results of the previous activity  have been reported and discussed 
in Ref.~\citenum{Frontera06}.
In short, the activity has mainly concerned a theoretical design study to establish the best 
design of a Laue lens telescope \cite{Pisa04,Pisa05a}, Monte Carlo
simulations of the expected optical properties of Laue lenses \cite{Pisa05b}, 
reflectivity measurements of mosaic crystal samples of Cu[111] \cite{Pellicciotta06}.
We have investigated the geometry of the lens, the crystal material
and lattice configuration that optimize the crystal reflectivity and energy passband.
Given that we have to cover with good  reflection efficiency a relatively broad 
energy  band (several hundreds of keV), special crystals, with properly controlled 
lattice deformations, appear to be more useful. Crystals of this kind
include mosaic crystals, bent crystals and crystals with non constant lattice 
spacing $d$ induced by doping materials or thermal gradients.
For our project we have assumed mosaic crystals, made of crystallites misaligned
each with other with controlled angular spread $\beta$ (FWHM of the Gaussian-like
angular distribution of the crystallite misalignments).  The growing technique of mosaic
crystals with the desired spread is now being consolidated (e.g., Courtois et al.\cite{Courtois04}). 

Crystal tiles of thickness $t$ are assumed to have their mean crystalline plane normal to 
the tile main facets, which are assumed to be square of side $l$. 

To correctly focus photons, the direction of the 
vector perpendicular to the mean lattice plane of each crystal has to intersect 
the lens axis, while its inclination with respect to the focal plane has to be equal 
to the Bragg angle $\theta_B$ (see figure in Ref.~\citenum{Pellicciotta06}).
The angle $\theta_B$ depends on the distance $r$ of the tile center from the lens axis 
and on the focal length $f$. For a correct focusing, it is needed that
$\theta_B = 1/2 \arctan{(r/ f)}$.
Once the crystal material is established, the Bragg angle increases with $r/f$, while
the energy of the focused photons decreases with $r/f$.
More generally, once the focal length is established, the outer  and inner
lens radii, $r_{max}$ and $r_{min}$, depend on the
nominal energy passband of the lens ($E_{min}$, $E_{max}$) and on the crystal 
lattice spacing: higher $d_{hkl}$ implies lower radii \cite{Frontera06}. For given
crystal material, outer lens radius and the focal length, the minimum energy that can be focused is 
established.

Thus, for a fixed inner and outer radius, the lens passband  can
be established by  the use of a combination of different crystal materials.
Among the candidate materials for their high reflectivity and for which the mosaic
technology has been developed (see, e.g., Ref.~\citenum{Courtois04}), Cu[111]
appears very promising for the hard X-/gamma--ray range \cite{Frontera06}. However,
for a long focal length (100 m), like in the case of the GRI mission, given the low
lattice spacing value $d_{111} = 2.087$~\AA\ of Cu[111], the minimum photon energy that can 
be focused is 320 keV for an outer lens radius of 185 cm (that of GRI lens).
The extension of the Laue lens passband below 320 keV can be obtained by the use of other 
crystal materials, e.g., Ge[111] with mosaic structure ($d_{111} = 3.266$~\AA) or
Si$_{1-x}$Ge$_x$[111] (Silicon $d_{111} = 3.135$~\AA) with a composition--gradient  
\cite{Abrosimov05}.
 
Mosaic spread and, for a fixed material, crystal thickness are the most crucial parameters
for an optimization of the lens performance. A single crystal thickness is not the 
best solution for optimizing the lens effective area in its entire passband.
However the optimization of the lens effective area at the highest energies could imply 
large thicknesses, that could be incompatible with lens weight constraints. 
We have investigated this issue \cite{Pisa05b}, finding that a good compromize
between crystal thickness and lens weight can be found.
  
Also the mosaic spread $\beta$ issue has been investigated. A higher spread gives a 
larger effective area, but also produces a larger defocusing of the reflected photons 
in the focal plane. By introducing a focusing factor $G$
\begin{equation}
G = f_{ph} \frac{A_{eff}}{A_d}
\end{equation}
 in which $A_{eff}$ is the effective area of the lens and $A_d$ is the
area of the focal spot which contains a fraction $f_{ph}$ of photons reflected by the lens,
it is found that, for long focal lengths like in the case of GRI (100 m), 
for its maximization, a very low spread ($\sim$30 arcsec) is requested.
However a lower  spread requires a higher accuracy in the positioning of the crystals 
in the lens. A compromize has to be found.

Another issue we have investigated is the disposition of the mosaic crystal tiles
in the lens for a uniform effective area of the lens passband. The best crystal tile 
disposition is an Archimedes' spiral that provides a smooth behavior of the lens 
effective area $A_{eff}$ with energy. However the Archimedes' spiral
becomes less important for long focal lengths ($>$ 30 m) and other approaches are
needed.

Also the required accuracy of the crystal tile positioning in the lens has been
investigated \cite{Pisa04}. It depends not only on the mosaic spread but also 
on the focal length. Higher focal lengths require higher positioning accuracies, 
which at the current stage of development is one of the major problems to be faced for
the realization of a Laue lens.

Results from a Monte Carlo (MC) code, that has been developed to derive the  
properties of different lens configurations, like their Point Spread Functions (PSF), 
for either on--axis and off--axis incident photons, have been already reported 
\cite{Pisa05a,Pisa05b,Frontera06}. They confirm 
the results obtained from the theoretical investigation and extend them. 

\begin{figure}
\begin{center}
\includegraphics[angle=0, width=0.4\textwidth]{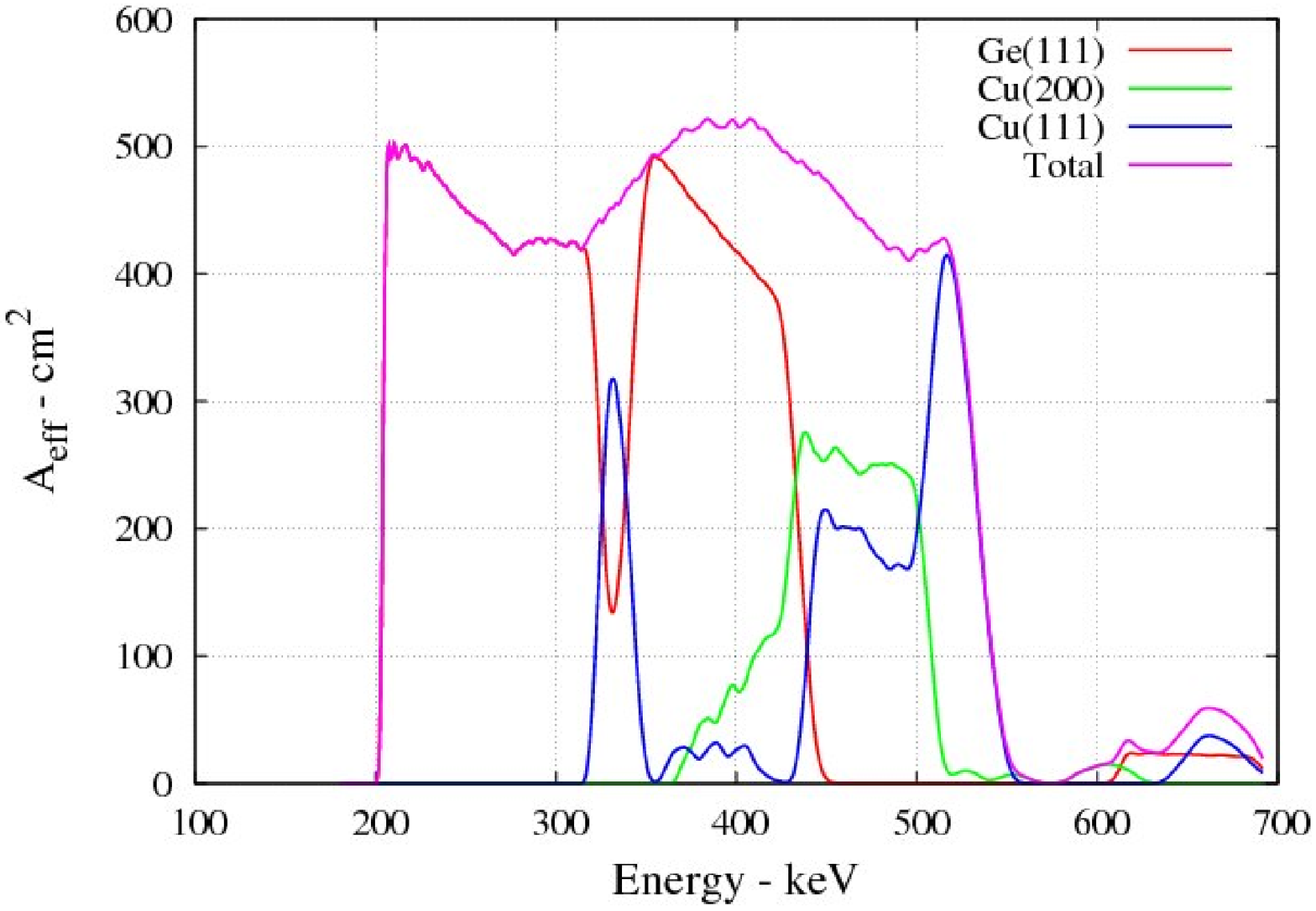}
\includegraphics[angle=0, width=0.4\textwidth]{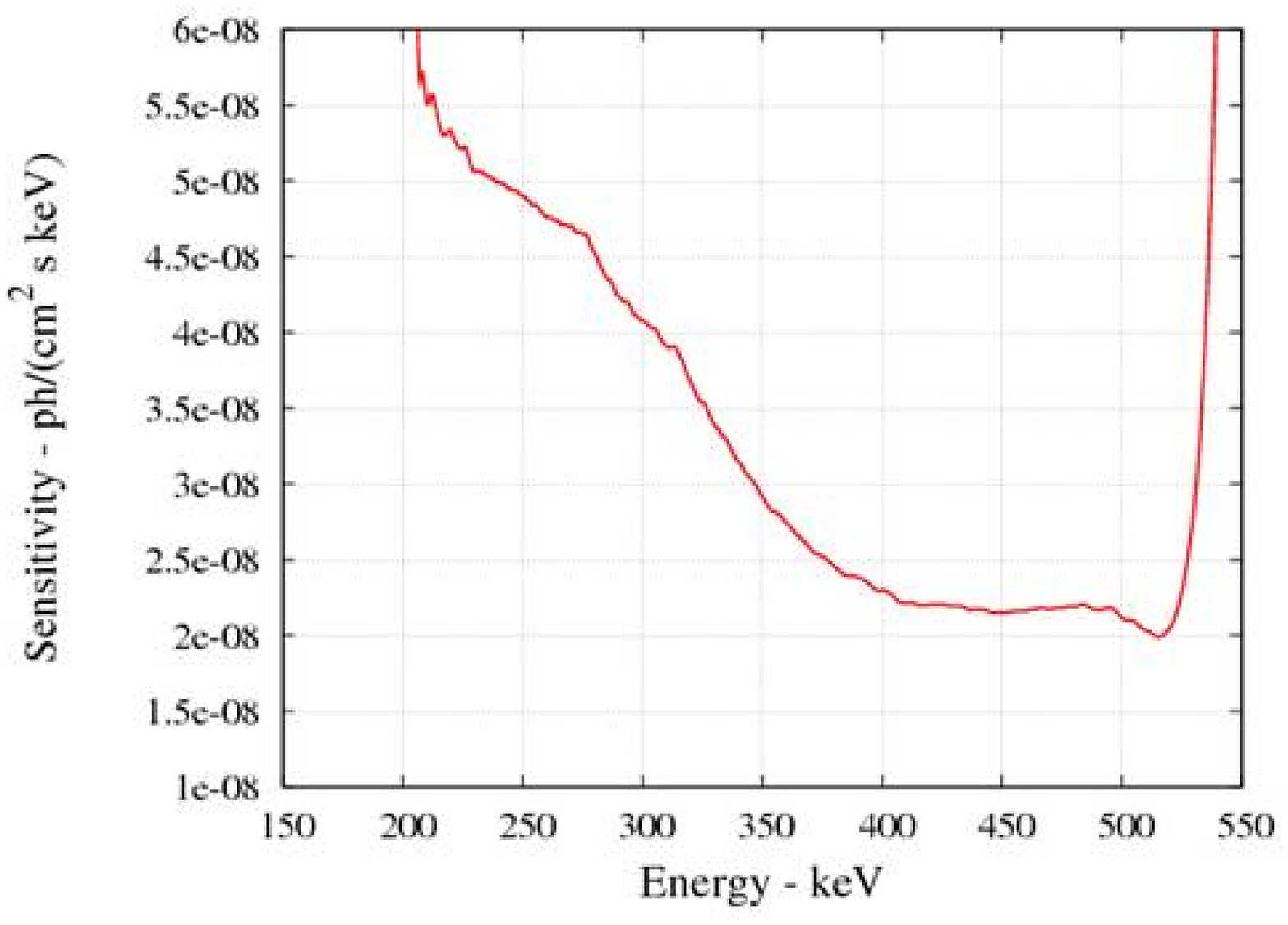}
\end{center}
\caption{Properties of the Laue lens described in Table~\ref{t:lens} in the 200--530 keV band
for an on-axis source(see text).  {\em Left panel:} On-axis effective area.
{\em Right panel:} the corresponding 3$\sigma$ sensitivity for an observation time
of $10^6$~s and energy channel width $\Delta E = E/2$.}
\label{f:onaxis_effarea_sens}
\end{figure}

We show in Fig.~\ref{f:onaxis_effarea_sens} the expected on--axis
effective area and 3$\sigma$  sensitivity of the lens described 
in Table~\ref{t:lens}. 
This lens configuration nicely fits also the GRI requirements.

The  expected  angular resolution of the lens is better than 1 arcmin.  

\begin{table}
\begin{center}
\caption{Main features of the simulated Laue lens.}
\label{t:lens}    
\begin{tabular}{llllll}
\hline\noalign{\smallskip}
Parameter &  \\	
\noalign{\smallskip}\hline \hline \noalign{\smallskip} \\
Focal length (m) &  100  \\ 
Nominal passband (keV) & 200--530 \\
Inner radius (cm) & 88 \\
Outer radius (cm) & 185 \\
Crystal material &  Ge[111], Cu[111], Cu[200] \\
Mosaic spread (arcmin) & 0.5 \\
Tile cross section (mm$^2$) & $15 \times 15$ \\
Tile thickness (mm)    &  optimized  \\
Number of crystal rings & 61 \\
No. of tiles & 17661 (Ge[111]), 3254 (Cu[200]), 3386 (Cu[111]) \\  
Crystal weight (kg)  & 155 \\
Effective area (cm$^2$) @ 200 keV &  500  \\ 
Effective area (cm$^2$) @ 400 keV &  530  \\ 
Effective area (cm$^2$) @ 511 keV &  430   \\
Half power radius(mm) &  12 \\
\noalign{\smallskip}
\hline
\end{tabular}
\end{center}
\end{table}

%
%

Results of reflectivity measurements of Cu[111], discussed elsewhere 
\cite{Pellicciotta06,Frontera05b}, confirm our expectations.

\section{Demonstration Model Development}

\begin{figure}
\begin{center}
\includegraphics[angle=0, width=0.4\textwidth]{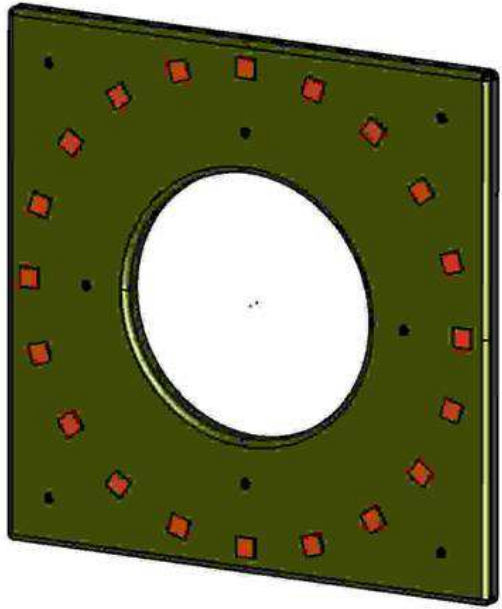}
\end{center}
\caption{Drawing of the lens Demonstration Model.}
\label{f:DM}
\end{figure}

A lens Demonstration Model (DM) is being developed. Unlike the Laue lens 
in Ref.~\citenum{Vonballmoos04}, the lens under development will
be made of crystal tiles rigidly fixed to the lens frame, without mechanisms for 
adjustment of their orientation. Thus their positioning in the lens
has to be correctly performed during the lens assembling. 
The goal of the DM development is just to establish the best crystal 
assembling technique of the lens. 
Figure~\ref{f:DM} shows the drawing of the DM under development.
%
%

It is composed of a ring of 20 mosaic crystals with diameter of 36 cm. The tiles are
made of Cu[111] with $\sim 3$ arcmin spread, 15$\times$15 mm$^2$ front surface and 
2 mm thickness.

\subsection{Lens assembly technique}

\begin{figure}
\begin{center}
\includegraphics[angle=0, width=0.4\textwidth]{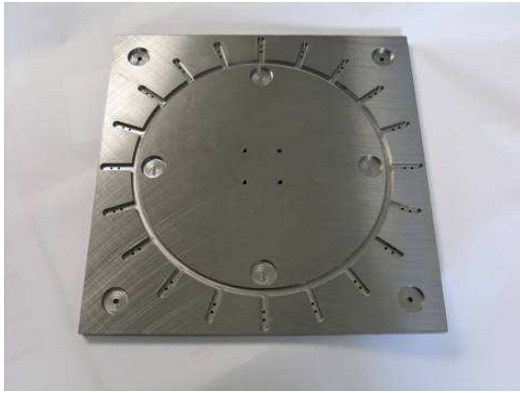}
\end{center}
\caption{DM counter-mask. The counter-mask is made of stainless steal. The holes,
2 per each crystal tile, are apparent.}
\label{f:countermask}
\end{figure}

The adopted lens assembling technique is based on the use of a counter-mask 
(see Fig.~\ref{f:countermask}). The counter-mask is provided with holes, two for each
crystal, with their axes directed toward the center of curvature of the
lens. Actually, in the case of the DM under development, the hole axis is parallel to that
of the lens axis. This simplification has been adopted in this phase
in order to focuse gamma--rays coming from a source at finite distance $d$ 
from the lens. In our case  $d \sim 6$~m.
%
%

\begin{figure}
\begin{center}
\includegraphics[angle=0, width=0.4\textwidth]{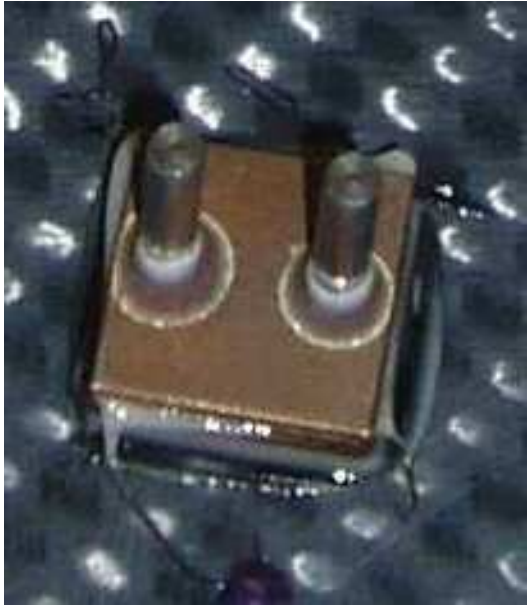}
\end{center}
\caption{Configuration of a crystal provided with 2 pins.}
\label{f:pins}
\end{figure}

Each crystal tile is positioned on the 
counter-mask by means of two pins (see Fig.~\ref{f:pins}) that are glued to each crystal with their 
axis parallel to the direction of the average crystalline plane. 

%
%

Once each crystal is positioned on the counter-mask, the lens frame is glued to all
the crystals. The lens frame is made of carbon fiber and is obtained starting from a  
mould. The lens is then separated from the counter-mask
by means of a chemical treatment that dissolves an aluminum cup that covers the pin base
closest to the crystal.
 
\begin{figure}
\begin{center}
\includegraphics[angle=0, width=0.4\textwidth]{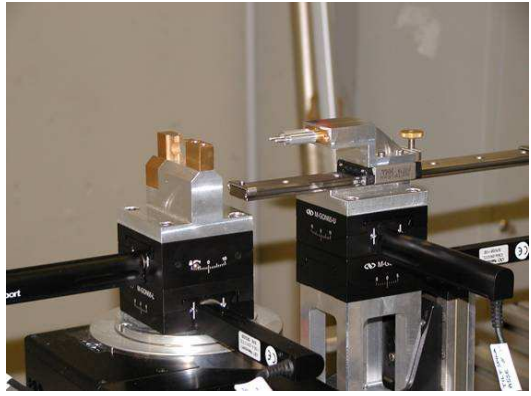}
\end{center}
\caption{A view of the set up for the crystalline plane determination and control
 and for the pin positioner.}
\label{f:pin_crystal_setup}
\end{figure}

Thus the assembling of the DM lens comprises the following main activities: 
\begin{itemize}
\item Proper alignment of each crystal on the X--ray optical bench;
\item Mounting of two support pins and bonding of pins to each crystal on the bench 
(see Figs.~\ref{f:pins} and \ref{f:pin_crystal_setup});
\item Positioning of all crystals (with support pins) on the counter-mask;
\item Bonding of all crystals to the lens frame;
\item Chemical etching of the support pins, and removal of the counter-mask.
\end{itemize}

%
%
%

\subsection{Test facility}

The set up for the crystalline plane determination and for the alignment
of the two pins with the crystalline plane direction, along with the gluing system
of the pins  to the crystal tiles, is shown in Fig.~\ref{f:diffr_transm_beams}. 
%
%

It is located in the X--ray facility LARIX (LARrge Italian
X--ray facility) of the University of Ferrara (for a description see
Ref.~\citenum{Loffredo05}). The gluing is performed when the
crystal correctly reflects the X--ray beam toward the lens focus. 
 
\begin{figure}
\begin{center}
\includegraphics[angle=0, width=0.4\textwidth]{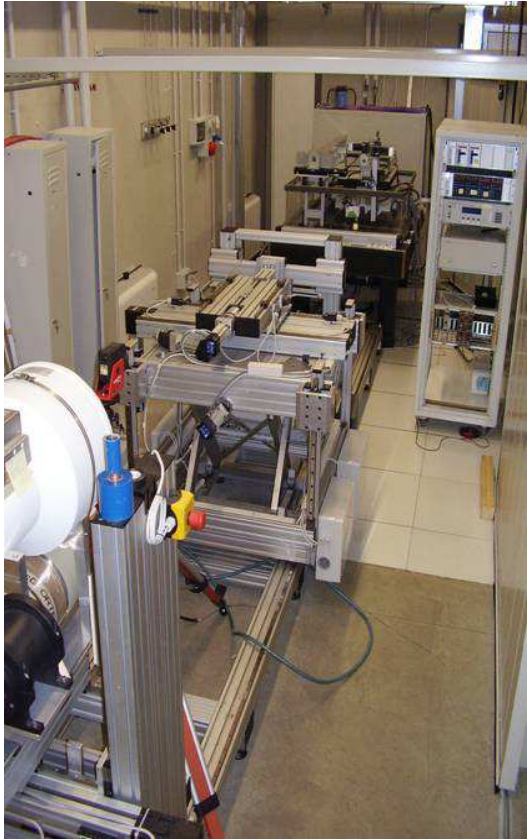}
\end{center}
\caption{A view of the apparatus for assembling  the
lens DM. The apparatus is located in the LARIX facility of the University
of Ferrara. }
\label{f:facility}
\end{figure}

\begin{figure}
\begin{center}
\includegraphics[angle=0, width=0.4\textwidth]{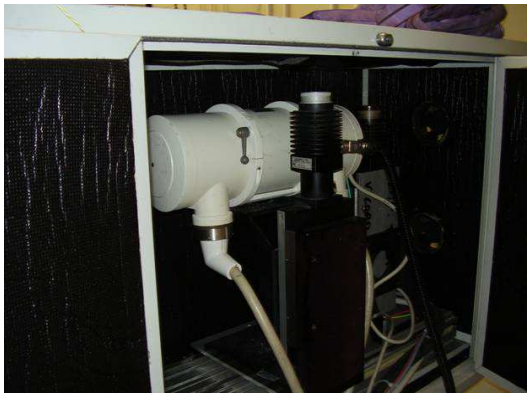}
\end{center}
\caption{A view of the X--ray generator tube, located in a lead box. }
\label{f:xray_tube}
\end{figure}

\begin{figure}
\begin{center}
\includegraphics[angle=0, width=0.4\textwidth]{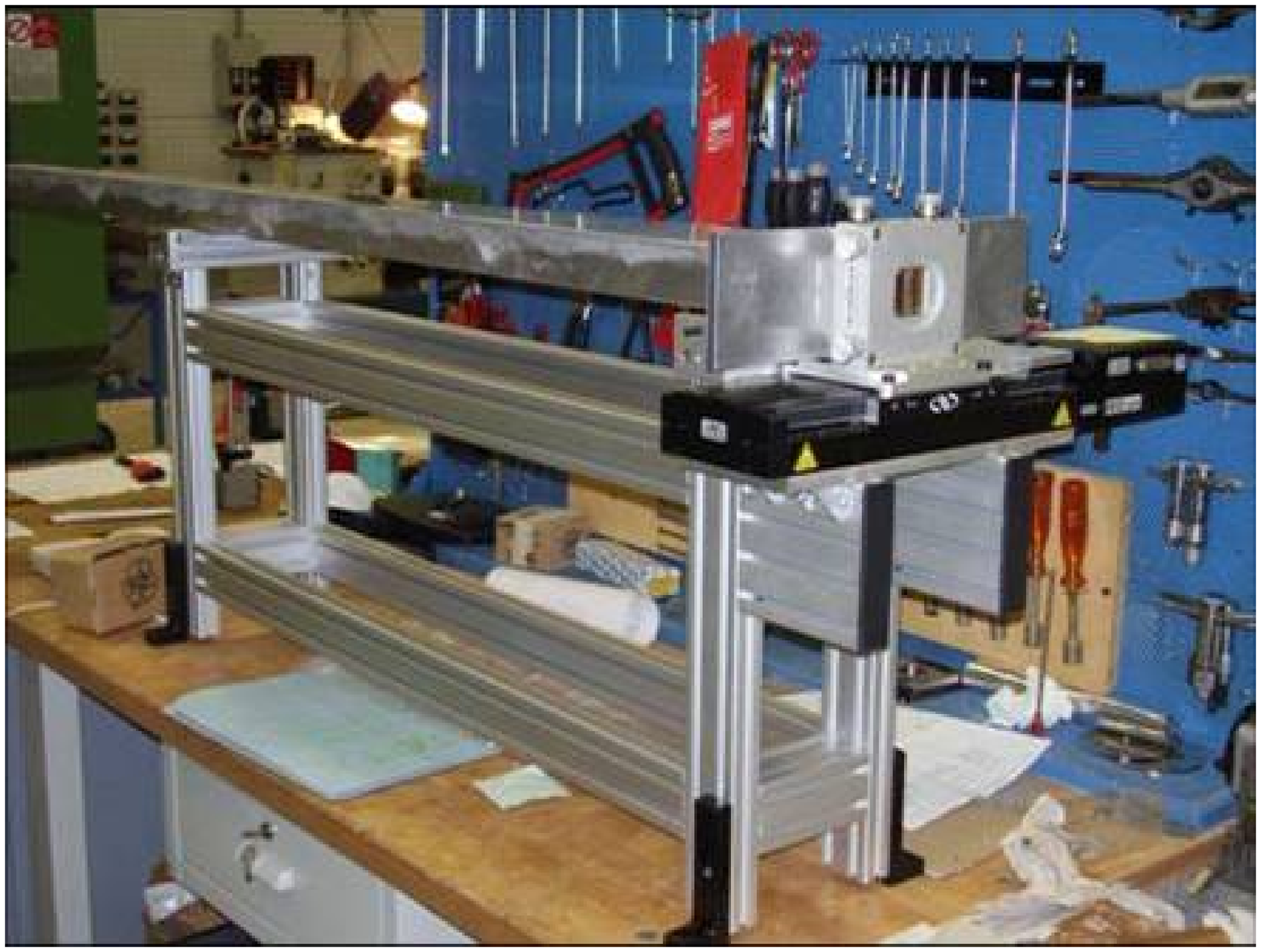}
\end{center}
\caption{A view of the pyramidal collimator during its assembling phase. In front 
of the collimator a movable Tantalum slit is visible.}
\label{f:collimator}
\end{figure}

A view of the experimental apparatus for assembling and testing the lens DM 
is shown in Fig.~\ref{f:facility}. 
%
%
The apparatus includes an X--ray generator tube (see Fig.~\ref{f:xray_tube})
 with a Tungsten target, a fine focus of 0.4 mm radius, a  maximum voltage of 150~kV and a 
maximum power of $\sim 200$~W. The X--ray tube, mounted on (X,Z) translation
stage, is located in a lead box in which a hole is made in correspondence of the X--ray. 
%
%
The X--ray photons coming out from the hole are collimated by a  pyramidal
collimator, at the end of which it is mounted a Tantalum slit with selectable
aperture along the vertical and the horizontal directions, both perpendicular to 
the pyramid axis. 
The slit can be also translated perpendicularly to the pyramid axis (see 
Fig.~\ref{f:collimator}).
The slit defines the cross section of the X--ray beam that irradiates 
the crystal main surface.   
%
%

\begin{figure}
\begin{center}
\includegraphics[angle=0, width=0.4\textwidth]{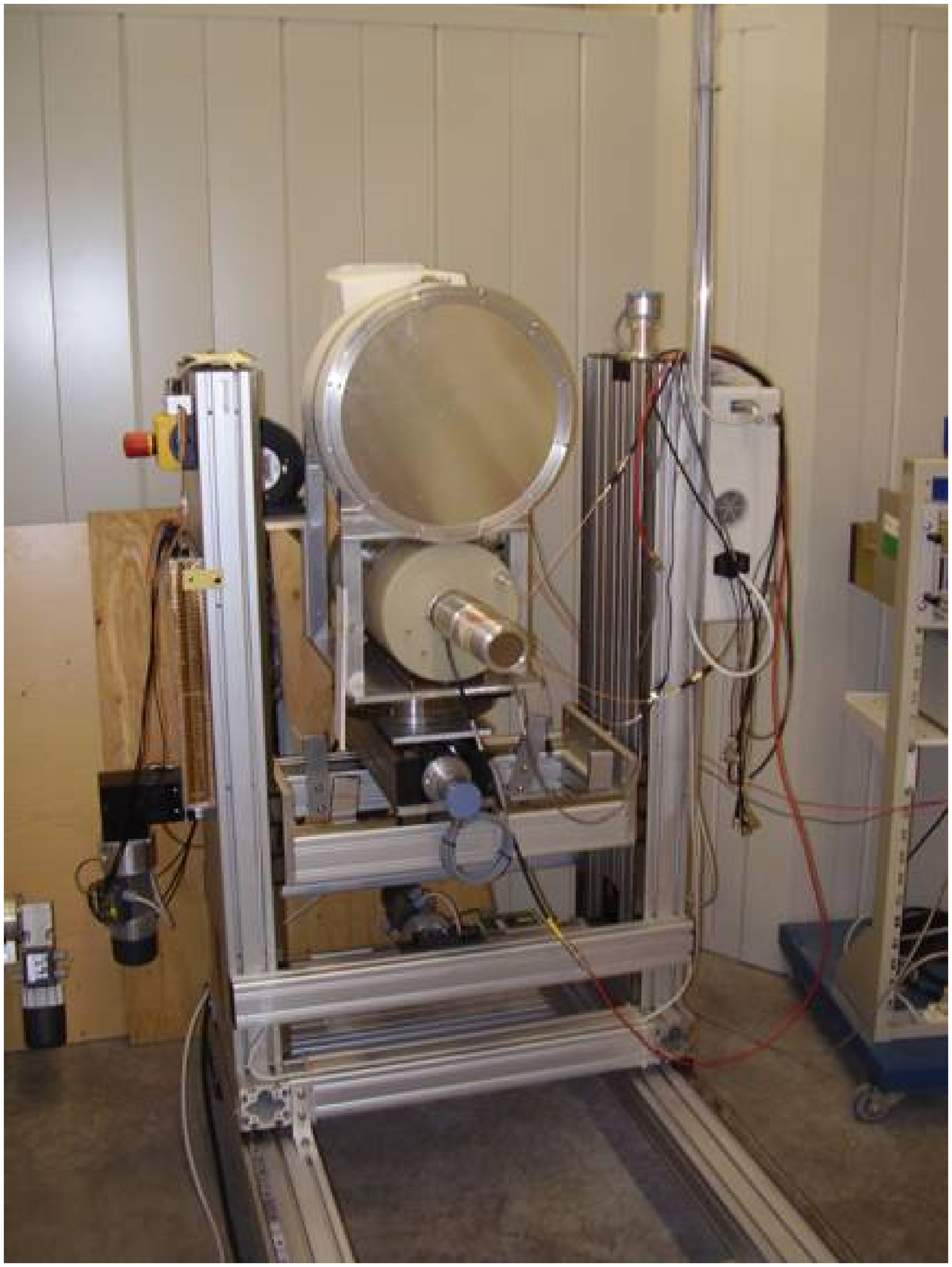}
\end{center}
\caption{A view of the detector set. The largest detector is the X--ray imager.}
\label{f:detectors}
\end{figure}

The radiation coming out from the slit collimator is used to both determine the
mean lattice crystalline plane of the crystal tile and to align the pins
to the X--ray beam. The image and spectrum of the
X--ray beam, either the direct one or that transmitted through the crystal tile or that
diffracted, can be detected. The available detectors include an X--ray imager,
a cooled HPGe detector and a position sensitive scintillator detector 
(see Fig.~\ref{f:detectors}). All of them can be rotated along the horizontal axis, and
can be translated along the vertical and along the horizontal. 
The position resolution of the X--ray imager is 0.3 mm.

%
%

\subsection{Determination of the average crystalline plane}

\begin{figure}
\begin{center}
\includegraphics[angle=0, width=0.3\textwidth]{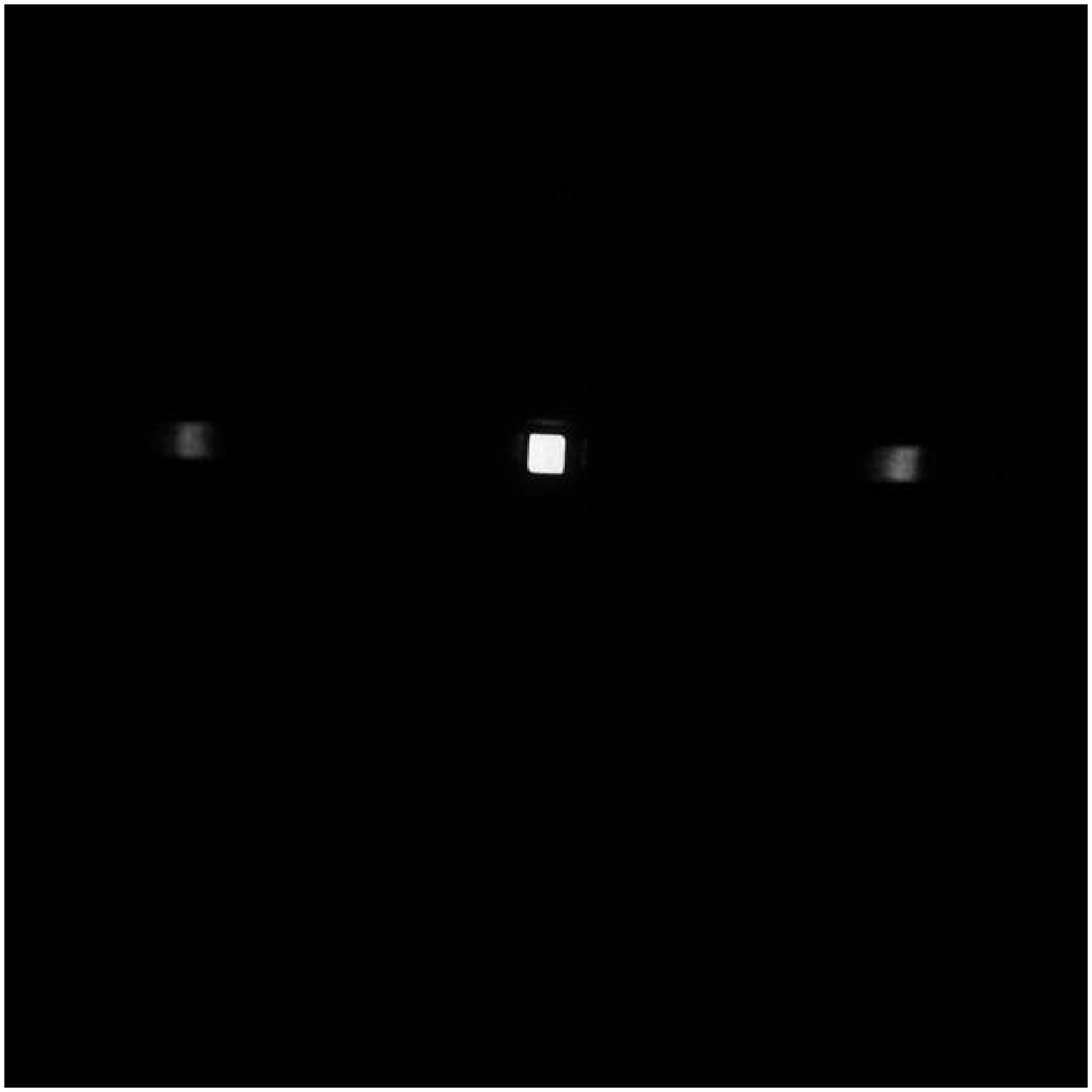}
\includegraphics[angle=0, width=0.5\textwidth]{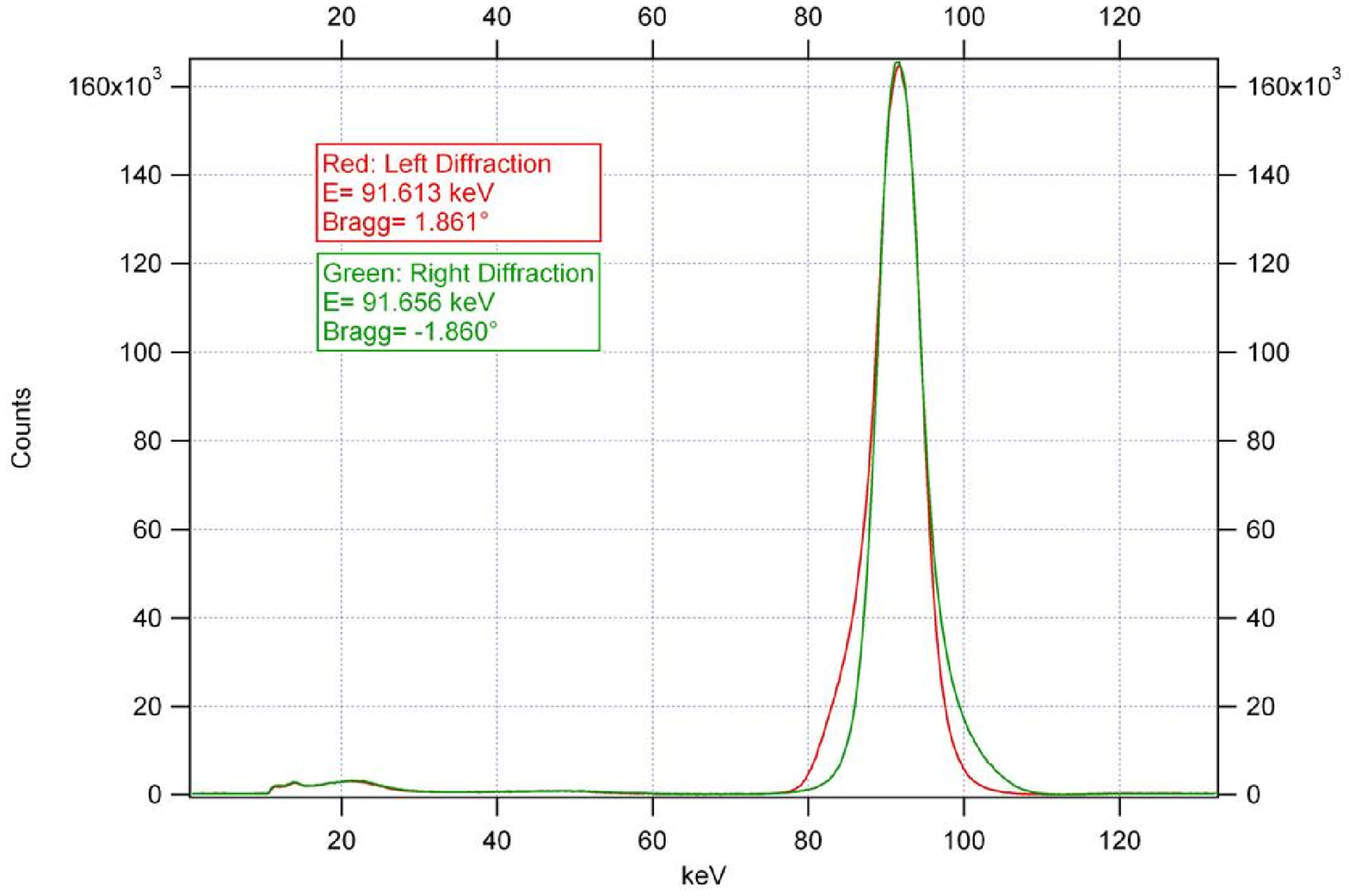}
\end{center}
\caption{{\em Left panel}: Superimposed images of the transmitted and diffracted
X--ray beam, once the crystalline plane is  found. {\em Right panel}: 
Diffracted spectra once the crystalline plane is found.}
\label{f:diffr_transm_beams}
\end{figure}

In order to determine the direction of the average lattice plane of the mosaic crystal tiles 
using the set up shown in Fig.~\ref{f:pin_crystal_setup},
the crystal tile is located on a crystal positioner. This can be
rotated around a vertical axis and  tilted along two orthogonal axes until
the diffracted beam from two symmetrical orientations of the crystal tile
give symmetrical images and coincident spectra (see 
Fig.~\ref{f:diffr_transm_beams}). An accuracy better than 10 arcsec in the determination
of the average crystalline plane is achieved.

%
%
\subsection{Alignment of the pin axis to the average crystalline planes}

\begin{figure}
\begin{center}
\includegraphics[angle=0, width=0.4\textwidth]{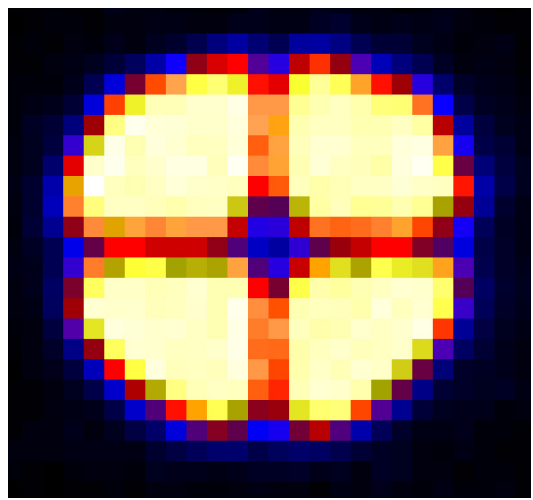}
\end{center}
\caption{Image of the X--ray shadow projected by the two crosses 
when their axes are almost aligned to the X--ray beam.}
\label{f:cross_on-image}
\end{figure}

The 2 pins needed for each crystal are mounted on the pin positioner (see
Fig.~\ref{f:pin_crystal_setup}).
This can be rotated along a circle with its center located in the  
vertical axis of the crystal positioner, and can be tilted along two 
orthogonal directions like the crystal positioner. The pin axis is requested
to be parallel to the beam axis, previously made parallel to the average crystalline
plane. This alignment is obtained by using 
X--ray shadow projected by two Tungsten crosses located along the pin positioner, with their
axes parallel to the pin axes (see Fig.~\ref{f:cross_on-image}).
%
%

\section{DM assembling status}

A preliminar DM model with few mosaic crystal tiles of Cu[111] with $\sim 3.5$~arcmin 
spread is expected to be assembled and tested in a short time to evaluate 
the cumulative error budget of the assembling technique. Soon after, the DM model made of a 
ring of Cu[111] mosaic crystal tiles will be assembled. Results of the first rigid Laue lens are
expected in a few month time.

\acknowledgments     
 
We acknowledge the financial support by the Italian Space Agency ASI and a minor contribution
by the Italian Institute of Astrophysics (INAF). The design study was also possible 
thanks to the received Descartes Prize 2002 of the European Committee.



\bibliography{lens_biblio}   
\bibliographystyle{spiebib}   

\end{document}